# Photo-fragmentation spectroscopy of benzylium and 1-phenylethyl cations.

Géraldine Féraud[a], Claude Dedonder-Lardeux[a], Satchin Soorkia[b] and Christophe Jouvet[a]

[a]Physique des Interactions Ioniques et Moléculaires, UMR CNRS 7345, Aix-Marseille Université, Avenue Escadrille Normandie-Niémen, 13397 Marseille Cedex 20, France

[b]Institut des Sciences Moléculaires d'Orsay, CNRS UMR 8214, Université Paris Sud 11, 91405 Orsay Cedex, France

Corresponding Author: christophe.jouvet@univ-amu.fr

The electronic spectra of cold benzylium ($C_6H_5\text{-}CH_2^+$) and 1-phenylethyl ($C_6H_5\text{-}CH\text{-}CH_3^+$) cations have been recorded via photofragment spectroscopy. Benzylium and 1-phenylethyl cations produced from electrosprayed benzylamine and phenylethylamine solutions, respectively, were stored in a cryogenically cooled quadrupole ion trap and photodissociated by an OPO laser, scanned in parts of the UV and visible regions (600-225 nm). The electronic states and active vibrational modes of the benzylium and 1-phenylethyl cations as well as those of their tropylium or methyl tropylium isomers have been calculated with *ab initio* methods for comparison with the spectra observed. Sharp vibrational progressions are observed in the visible region while the absorption features are much broader in the UV. The visible spectrum of the benzylium cation is similar to that obtained in an argon tagging experiment [V. Dryza, N. Chalyavi, J.A. Sanelli, and E.J. Bieske, J. Chem. Phys. **137**, 204304 (2012)], with an additional splitting assigned to Fermi resonances. The visible spectrum of the 1-phenylethyl cation also shows vibrational progressions. For both cations, the second electronic transition is observed in the UV, around 33 000 cm$^{-1}$ (4.1 eV), and shows a broadened vibrational progression. In both cases the $S_2$ optimized geometry is non planar. The third electronic transition observed around 40 000 cm$^{-1}$ (5.0 eV) is even broader with no apparent vibrational structures, which is indicative of either a fast non-radiative process or a very large change in geometry between the excited and the ground states. The oscillator strengths calculated for tropylium and methyl tropylium are weak. Therefore, these isomeric structures are most likely not responsible for these absorption features. Finally, the fragmentation pattern changes in the second and third electronic states: $C_2H_2$ loss becomes predominant at higher excitation energies, for both cations.

# I. Introduction

Carbocations are ions with a positively charged carbon atom.[1] In particular, aromatic ions appear frequently as short-lived reactive intermediates in organic reaction mechanisms.[2] There has been considerable interest in the study of carbocations because of their importance in chemical reactions.[3–5] The knowledge of their structures and properties is central to the understanding of their chemical reactivity, and IR spectroscopy methods have contributed much to a better understanding of the ground state of these species.[6,7] However, very little information is available on their electronic spectroscopy in the literature.

While electronic transitions of neutral aromatic molecules are more energetic (UV range), the lowest-lying electronic states of aromatic cations are in the visible region.[8–10] Because of their large energy range of electronic transitions and their associated width, aromatic ionic species are regarded as potential carriers of the Diffuse Interstellar bands (DIBs).[11–14]

Some closed-shell aromatic cations have been characterized in rare gas matrices[15,16] and in the gas phase.[7,17–21] Protonated species can be produced by several techniques such as chemical ionization,[17] pulsed-electrical discharge[20] or in an electrospray ionization source (ESI),[18,19] and investigated through IR multiple photon dissociation spectroscopy (IRMPD)[7,17,19] or cold ion UV photofragment spectroscopy.[22,23] Cold traps that can be cooled down to ~10 K have become the ideal tool for studying optical properties of isolated gas phase bio-molecules, like protonated amino acids and peptides.[18,24] The ESI source is a very convenient ion source: not only it allows to produce protonated molecules, but also radical cations and collision induced fragments of protonated molecules,[25–31] which can further be trapped and cooled down. Therefore ESI is an easy technique to produce gas phase radical or closed-shell cations, and the coupling with the low temperatures attainable in a cold ion trap greatly simplifies the study of the electronic spectroscopy of such large isolated molecular species.

The $C_7H_7^+$ ion has long been a challenge to mass spectroscopists with its two possible cyclic isomeric forms: the benzylium cation ($C_6H_5CH_2^+$), a six membered ring with an attached methylidene group, stabilized through resonance and the tropylium cation, a seven membered ring, which is extremely stable because of its six π electrons aromatic structure. It is known that these two isomers can undergo interconversion.[15,32–37] The benzylium (or tropylium) ion $C_7H_7^+$ (m/z 91) and the 1-phenylethyl cation $C_8H_9^+$ (m/z 105) are among the simplest aromatic closed-shell cations and specifically prominent in the mass spectra of alkylbenzenes and related compounds.[33–35,37–40] For the $C_8H_9^+$ ion, rearrangement pathways between the 1-phenylethyl cation and the methyl tropylium cation have also been discussed.[32,41,42]

Until recently, only broad electronic absorptions have been observed for $C_7H_7^+$ cations in the solution phase, in Argon matrices, and in ion traps with significant differences in band positions and assignments.[15,36,43–46] More recently, vibrationally resolved electronic spectra of benzylium and tropylium cations have been recorded in a Neon matrices,[16] but the influence of the matrix is still non negligible, i.e. band broadening and hindered rotational motions. The Argon-tagged benzylium cation has also been studied through laser-induced dissociation[47] and two electronic states have been evidenced, one in the visible and one in the UV region. The calculated electronic structure and simulated vibronic spectra are in very good agreement with the experiment.

From an experimental as well as theoretical point of view, the 1-phenylethyl cation has received less attention compared to the benzylium and tropylium cations. Recently, IRMPD spectroscopy has been used to differentiate between the 1-phenylethyl form and the methyl-tropylium structure[42] as well as to characterize hydroxy- substituted benzylium and tropylium ions.[48]

We present here the photo-fragmentation spectra of cold and isolated gas phase benzylium and 1-phenylethyl cations, on a very wide spectral range in the visible and UV regions. The assignment of electronic states and vibrational modes is achieved by comparison with *ab initio* calculations.

## II. Methods
### A. Experiment

The experimental setup, similar to the one developed by Wang and Wang,[18,38-39] has already been described.[23] Ions are produced in an electrospray ionization source (from Aarhus University), stored in a cryogenically cooled quadrupole ion trap (Paul trap from Jordan TOF Products, Inc.) where they interact with a laser beam. When the trapped ions absorb photons, they may fragment and the ionic photo-fragments are detected in a time-of-flight spectrometer ($M/\Delta M = 300$).

Benzylium and 1-phenylethyl cations are produced by collision-induced dissociation of electrosprayed protonated benzylamine and protonated phenylethylamine, respectively. Dissociation occurs in the high-pressure region of the ion source, between the capillary and the skimmer. At the exit of the capillary, ions are guided and trapped in an octopole trap and are further extracted and accelerated with pulsed voltages in order to produce ion packets with a duration between 500 ns and 1 µs. A mass gate placed at the entrance of the Paul trap allows selecting the ions of interest, i.e. benzylium (m/z 91) or 1-phenylethyl (m/z 105). The trap is mounted on a cold head of a liquid Helium cryostat (Coolpak Oerlikon). The cations are trapped and thermalized through collisions with the buffer gas (Helium). The temperature of the cold trap is monitored with two temperature sensors: the first sensor ($T_1$) is located on the cryostat head on which the trap is mounted and the second sensor ($T_2$) is on the top of the Paul trap. Typically, the temperatures attained are $T_1 = 14$ K and $T_2 = 40$ K. Therefore, the temperature of the trap is ~40 K or slightly lower. The spectral resolution of the OPO laser does not allow to record the rotational contour for an estimate of the vibrational temperature of the ions. But in the present case, hot bands that can be identified with calculations are very weak in intensity: they appear as shoulders in simulations with a temperature of 100 K, but do not appear on the experimental band origin. In the case of protonated phenylalanine cations, we recorded a spectrum showing a hot band, and the intensity ratio between the origin band and the hot band gave a temperature of c.a. 50 K. The ions are kept in the trap for several tens of ms (typically 10 to 90 ms) before the photo-dissociation laser is triggered: this delay is necessary to ensure thermalization of ions and evacuation of He buffer gas from the trap. Under these conditions, cations are not fragmented by collisions when they are extracted and accelerated for analysis in the linear time-of-flight mass spectrometer. Photo-fragmentation spectra are obtained by recording the fragment signal of cations on a microchannel plates (MCP) detector as a function of laser wavelength with a digitizing storage oscilloscope interfaced to a PC.

The photo-dissociation laser is an OPO laser from EKSPLA, which has a 10 Hz repetition rate, a resolution of 8 cm$^{-1}$ and a scanning step of 0.04 nm. The laser is mildly focused in the trap on a 3 mm diameter spot, the laser power being around 20 mW in the UV and 40 mW in the visible.

### B. *Calculations*

Ab initio calculations have been performed with the TURBOMOLE program package,[51] making use of the resolution-of-the-identity (RI) approximation for the evaluation of the electron-repulsion integrals.[52] The equilibrium geometry of ground electronic state ($S_0$) has been determined at the MP2[53] level. Excitation energies and equilibrium geometry of the lowest excited singlet state ($S_1$) have been determined at the RI-CC2 level.[43-44] Calculations were performed with two basis sets: the correlation-consistent polarized valence triple-zeta cc-pVTZ basis set for the benzylium cation and the double-zeta cc-pVDZ basis set for the 1-phenylethyl cation.[56]

The vibrations in the ground and excited states have been calculated and the vibronic spectra simulated using PGOPHER software[57] for the Frank-Condon analysis.

## III. Results
### A. Experiment
#### 1. Benzylium cation (m/z 91)

In this part, we present the photo-fragment spectroscopy of the benzylium cation in the visible (18 800 cm$^{-1}$ – 24 400 cm$^{-1}$) and the UV (30 000 cm$^{-1}$ – 44 500 cm$^{-1}$). The main ionic fragments issued from the photo-dissociation of benzylium ions (m/z 91) are (i) $C_5H_5^+$ ion (m/z 65) issued from acetylene loss (-$C_2H_2$) and (ii) $C_3H_3^+$ ion (m/z 39) a fragment frequently observed in the dissociation of aromatic ions corresponding to $C_4H_4$ loss (or two $C_2H_2$ loss) (see Figure SI1 in the Supplementary Material).

##### a. Visible spectrum

Figure 1 shows the photo-fragment spectrum of cold benzylium (Bz$^+$) cation in the visible, recorded on the m/z 65 fragment. It is a well-structured $S_1 \leftarrow S_0$ transition, with the band origin observed at 523.9 nm (2.36 eV, 19 089 cm$^{-1}$). This value is very close to the one extrapolated from the Bz$^+$-Ar and Bz$^+$-Ar$_2$ transition measurements by Dryza *et al.* (19 082 cm$^{-1}$).[47] The spectrum has been recorded further to the red, i.e. up to 800 cm$^{-1}$ from the band origin (see Figure SI2 in Supplementary Material), and no other bands have been detected to the red of the origin. Above 24 400 cm$^{-1}$, the first excited state photo-fragmentation spectrum decreases rapidly and no other absorption bands appear between 25 000 and 30 400 cm$^{-1}$.

The spectrum exhibits a long vibrational progression in $\nu_{13}$ = 504 cm$^{-1}$ (Mulliken notation). This progression has already been discussed elsewhere.[14,16,47] Apart from the 0-0 transition which is a single band, all other vibrational bands are split by c.a. ~10 cm$^{-1}$. This splitting, not observed in other experiments,[16,47] increases with the number of quanta (14 cm$^{-1}$ for one quantum in $\nu_{13}$ to 28 cm$^{-1}$ for 3 quanta in $\nu_{13}$). Our calculations, in agreement with previous ones,[47] indicate that there is only one vibrational mode around 500 cm$^{-1}$ and the splitting could be due to Fermi resonances. A combination band of two out-of-plane vibrations can be responsible for this splitting (vide infra).

In addition to the main progression, three other progressions starting at 1176 cm$^{-1}$, 1236 cm$^{-1}$ and 1350 cm$^{-1}$ above the band origin are observed (Fig. 1). The first frequency corresponds to $\nu_9$ mode (observed at 1181 cm$^{-1}$ in the experiment of Dryza *et al.*).[47] The other modes will be tentatively assigned on the basis of calculated frequencies in the discussion section. Above 24 400 cm$^{-1}$, the first excited state photo-fragmentation spectrum decreases rapidly and further absorption bands appear higher in energy at around 30 000 cm$^{-1}$ (see next section).

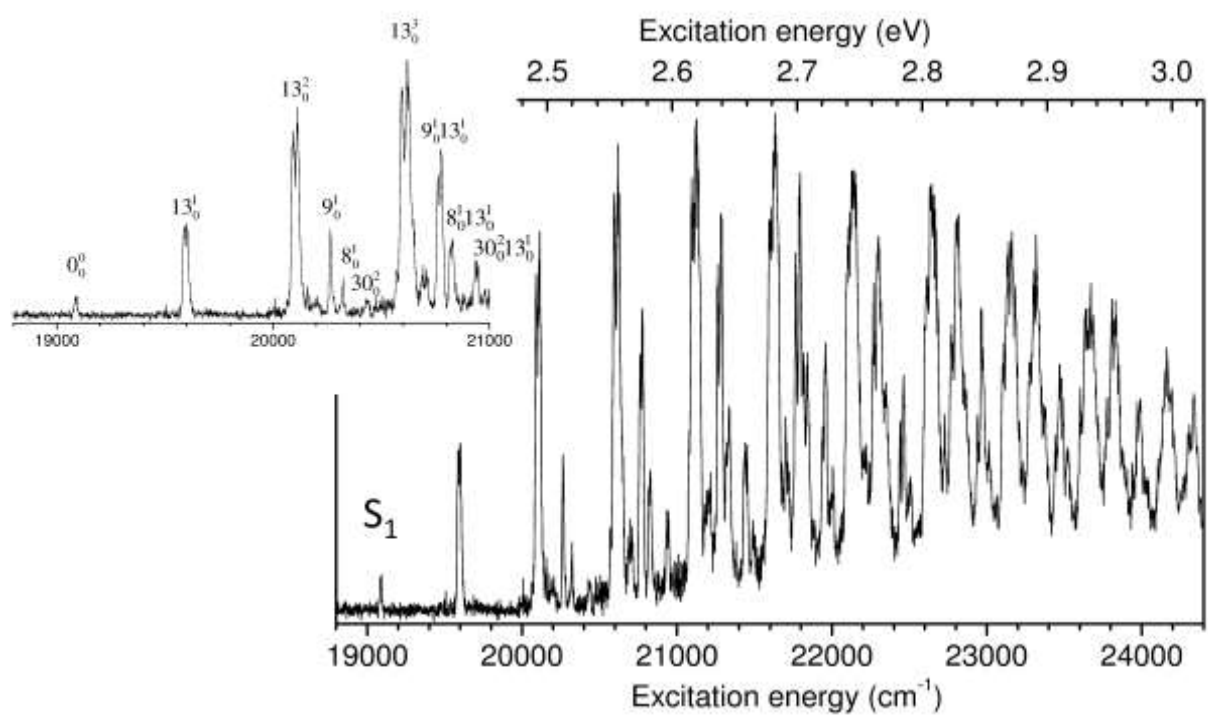

FIG. 1. Photofragmentation spectrum of the benzylium cation in the visible region (S$_1$ state), recorded on the m/z 65 fragment. In the inset, enlarged view of the first vibronic bands. The splittings are assigned to Fermi resonances.

*b. UV spectrum*

At higher energy, in the UV, two states have been identified: $S_2$ starting at about 32 000 cm$^{-1}$ (3.95 eV) and $S_3$ around 40 000 cm$^{-1}$ (5 eV) (Fig. 2). A broadened vibrational progression on the $\nu_{13}$ mode (c.a. 500 cm$^{-1}$) is observed for the $S_2$ state, and no apparent vibrational structure is observed for the $S_3$ state. The absence of well-structured vibronic bands is a clear indication that some fast dynamics occurs in these excited states, which smears out the vibrational structure.

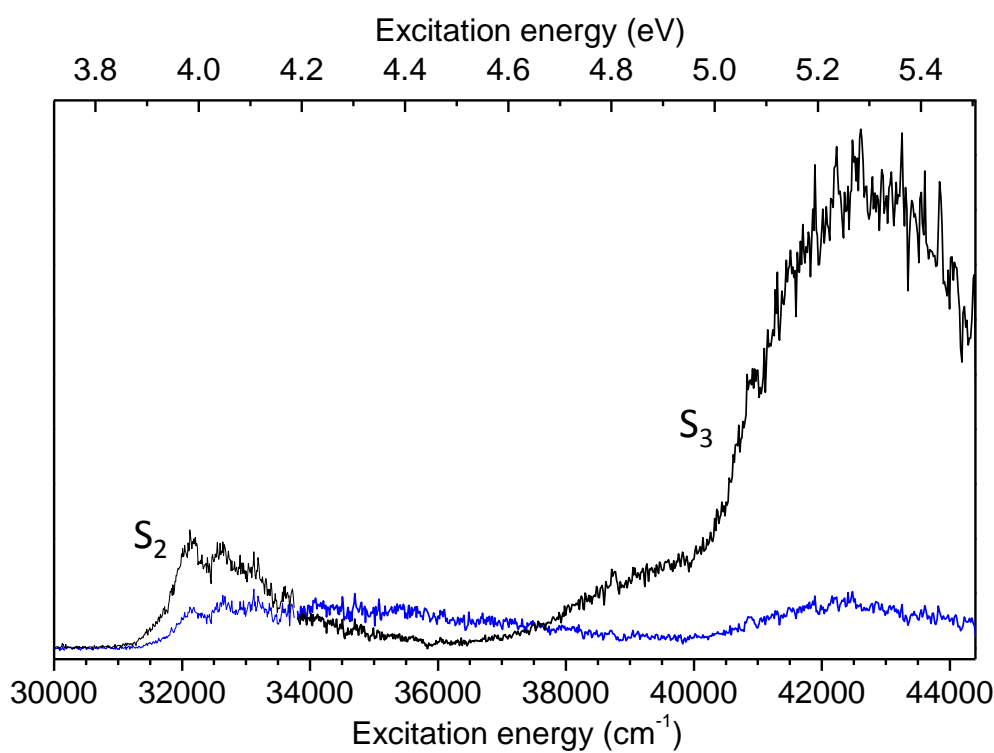

FIG. 2. Photofragmentation spectrum of the benzylium cation in the UV region, recorded on the m/z 39 (blue) and m/z 65 (black) fragments.

The branching ratio between m/z 39 and m/z 65 photofragments changes when the energy increases: at the onset of the $S_2$ state at 32 000 cm$^{-1}$, m/z 65 is stronger in intensity than m/z 39 but decreases within 4000 cm$^{-1}$ further in the blue, while in the S$_3$ state, at higher energy, m/z 65 (C$_2$H$_2$ loss) is predominant.

### 2. 1-phenylethyl cation (m/z 105)

*a. Visible spectrum*

The photo-dissociation spectrum of the C$_6$H$_5$-CH-CH$_3^+$ 1-phenylethyl cation (m/z 105) in the visible region (20 700 cm$^{-1}$ – 22 500 cm$^{-1}$) is presented in Figure 3. The ionic fragments of 1-phenylethyl cation are m/z 77 (C$_2$H$_2$+H$_2$ or C$_2$H$_4$ loss), m/z 79 (C$_2$H$_2$ loss) and m/z 103 (H$_2$ loss) in the 3 : 4 : 2 ratio. They all show the same spectrum, and therefore the signals from all the fragment ions have been averaged.

The first band of the electronic spectrum of the 1-phenylethyl cation recorded at 21 068 cm$^{-1}$ is weak in intensity. We have recorded the spectrum up to 600 cm$^{-1}$ to the red of the first band (figure SI3 in the Supplementary Material) and no other bands have been observed. Therefore, the band at 21 068 cm$^{-1}$ is assigned to the band origin of the S$_1$←S$_0$ transition. The spectrum presents some similarities with the benzylium spectrum (inset of Fig. 3), exhibiting a progression on a vibrational mode of around 400 cm$^{-1}$, on which is superimposed an additional vibrational progression based on a low frequency mode around 50 cm$^{-1}$. This type of vibrational structure is reminiscent of the hindered rotation of a methyl group.[58] In the 24 400-28 600 cm$^{-1}$ region the first excited state band decreases to zero, without structures.

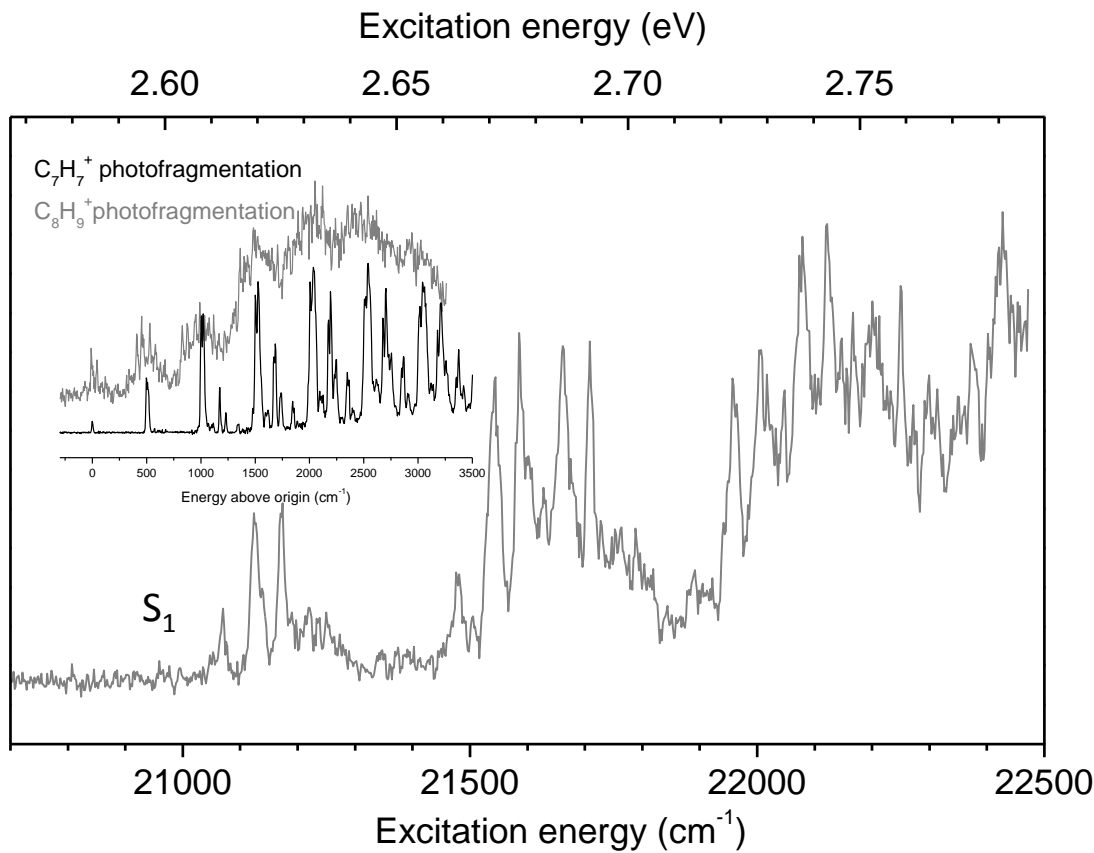

FIG. 3. Photo-fragmentation spectrum of the 1-phenylethyl cation in the visible region ($S_1$ state). This spectrum corresponds to the average of the signals recorded on the m/z 77, 79 and 103 photo-fragments. Inset: comparison between the visible spectra of $C_8H_9^+$ 1-phenylethyl (grey) and $C_7H_7^+$ benzylium cations (black).

b. *UV spectrum*

At higher energy (between 28 500 and 44 500 cm$^{-1}$), in the UV, the observed signal is quite complex and broad, as for the benzylium ion (Fig. 4). The first band starting at 29 500 cm$^{-1}$ (3.65 eV) exhibits a double bump structure on which is superimposed a broad progression similar to the one observed in the visible region (inset of Fig. 4). In the latter region, the fragmentation pattern does not change with energy while in the UV, the branching ratio changes with increasing energy: the production of the lightest fragment (m/z 77), which corresponds to $C_2H_4$ (or CH-CH$_3$) loss, decreases while the $C_2H_2$ loss (leading to m/z 79) is favored with increasing energy. The H$_2$ loss channel (m/z 103) is also present and has the same spectral pattern as m/z 79.

Further in the blue, a broad band with no apparent vibrational structure is observed, centered at 39 000 cm$^{-1}$ (4.8 eV), leading predominantly to C$_2$H$_2$ loss, as was the case for the benzylium ion.

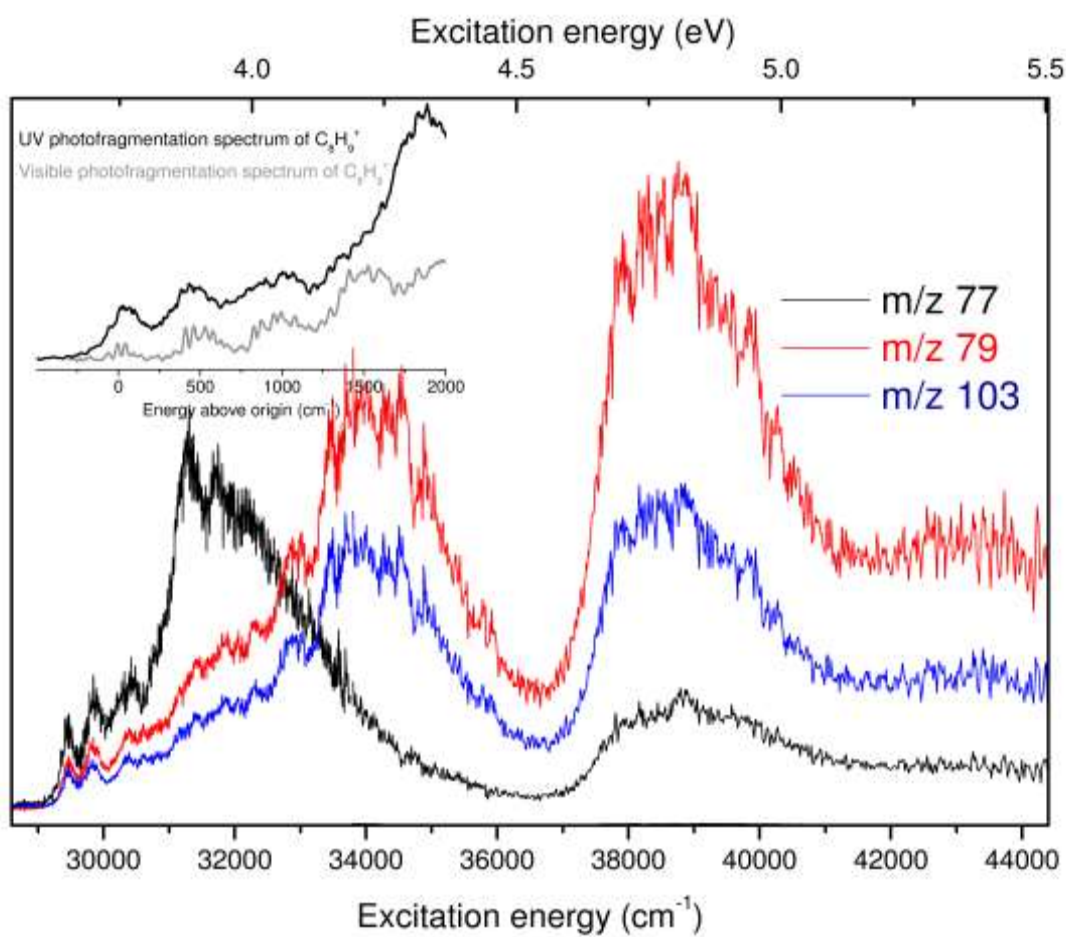

FIG. 4. Photofragment spectrum of 1-phenylethyl cation in the UV region, recorded on m/z 77 (black) and m/z 79 (red) and m/z 103 (blue). Inset: comparison between the S$_1$ ← S$_0$ in the visible region (grey) and S$_2$ ← S$_0$ first transition in the UV region of 1-phenylethyl cation. The m/z 103 (loss of H$_2$) and m/z 79 (C$_2$H$_2$ loss) show the same pattern over the whole energy region while the spectrum of the m/z 77 fragment (C$_2$H$_4$ loss) is different in the low energy region.

### B. Calculations

#### 1. Benzylium cation (m/z 91)

The benzylium cation has been calculated extensively in two recent papers[14,47] using different methods (TD-DFT mostly). It has a $C_{2v}$ structure in the ground state and the vertical energies and the oscillator strengths calculated at the RI-CC2/cc-pVTZ level are presented in Table I. The main difference between the CC2 and the DFT calculations[47] is the ordering of the $S_2$ and $S_3$ states.

The calculated adiabatic energy for the $S_1$ ($^1B_1$) state is 2.51 eV when the difference in zero-point energy, $\delta$ZPE, is taken into account. This value compares quite well with the experimental transition origin (2.36 eV). As in the previous studies,[14,47] the main excited state vibrational progression can be well reproduced using the active mode $\nu_{13}$ (calculated at 502 cm$^{-1}$) and three other progressions on mode $\nu_{13}$ in combination with $\nu_9$ (calculated at 1187 cm$^{-1}$ and observed at 1176 cm$^{-1}$), $\nu_8$ (observed at 1236 cm$^{-1}$ and calculated at 1203 cm$^{-1}$) and a fourth mode observed at 1350 cm$^{-1}$, which is tentatively assigned to 2 quanta in $\nu_{30}$ (calculated at 1346 cm$^{-1}$).

The $S_2$ state can be optimized when the $C_{2v}$ symmetry is maintained and the adiabatic transition energy is then calculated at 4.43 eV. When the symmetry constraint is released, the $CH_2$ group starts to rotate and the full optimization leads to a geometry in which the $CH_2$ plane is perpendicular to the aromatic ring, as calculated in TD-DFT.[47] In the optimization process, the $S_2$ state crosses $S_1$ at around 3 eV.

TABLE I. Calculated vertical and adiabatic (bold italic) excited state transition energies (eV) and oscillator strengths (italic in parenthesis) for the benzylium cation in comparison with previous studies and with the experimental transition origin (eV).

| $C_{2v}$ | Symmetry | Vertical/*Adiabatic* transition energy *(oscillator strength)* | | | | transition origin |
|---|---|---|---|---|---|---|
| | | RI-CC2/ | TD-DFT | TD-DFT | TD-DFT | exp. |

|  |  | cc-pVTZ[a] | BLYP/6-311G(d,p)[b] | B3LYP/cc-pVDZ[c] | M06/aug-cc-pVDZ[d] |  |
|---|---|---|---|---|---|---|
| $S_1$ | $^1B_1$ | 3.135/***2.67***[e] (0.035) | 2.73 (0.0193) | 3.022/***2.580*** (0.0258) | 3.02 (0.026) | 2.36 |
| $S_2$ | $^1A_1$ | 4.702/***4.43*** (0.41) | 4.42 (0.2257) | 4.687 (0.3164) |  | 3.95 |
| $S_3$ | $^1A_2$ | 5.033 (0.00) |  | 4.539 (0.0000) |  |  |
| $S_4$ | $^1B_2$ | 5.448 (0.000074) |  | 5.022 (0.0002) |  |  |
| $S_5$ | $^1A_1$ | 6.278 (0.099) |  |  |  |  |
| $S_6$ | $^1A_2$ and $^1B_1$ | 6.776 (0.004) |  |  |  |  |

[a]this work; [b]Nagy et al.;[16] [c]Dryza et al.;[47] [d]Troy et al.;[14] [e]2.51 eV with the difference in zero-point energy correction.

It is noteworthy to mention that the excited states calculated in the 5 eV region, where an intense band is observed (Fig. 2), have very low (or zero) oscillator strength.

### 2. Tropylium cation (m/z 91)

One could argue that some of the UV bands could be assigned to the tropylium ion, as in the neon matrix study.[16] Tropylium has a $D_{7h}$ symmetry. Since TURBOMOLE[51] does not handle CC2 calculations with $D_{7h}$ symmetry, this cation has been calculated using the TD-DFT (B3LYP/cc-pVDZ) method with the $D_{7h}$ symmetry and using the RI-CC2/cc-pVDZ method with a lower $C_{2v}$ symmetry. The lowest excited states are found at similar energies with the two methods. The results of excited state calculations in TD-DFT (B3LYP/cc-pVDZ) for the tropylium ion are presented in Table II.

TABLE II. Vertical transition energies (eV) and oscillator strengths for the excited states of the tropylium ion calculated in TD-DFT (B3LYP/cc-pVDZ). Values in parentheses are obtained with the RI-CC2/cc-pVDZ method in $C_{2v}$ symmetry.

| Symmetry | Vertical transition energy | Oscillator strength |
|---|---|---|
| $^1E'_3$ | 4.77 (4.81[a]) | 0 |
| $^1A''_1$ | 5.86 (6.37) | 0 |
| $^1E''_3$ | 5.88 | 0 |
| $^1A''_2$ | 5.94 (6.38) | 0.00387 |
| $^1E'_1$ | 6.18 | 1.20 |
| $^1E''_1$ | 7.08 | 0 |

| $^1E''_2$ | 7.19 | 0 |
|---|---|---|
| $^1E'_2$ | 7.21 | 0 |
| $^1A'_2$ | 12.65 | 0 |
| $^1A'_1$ | 12.95 | 0 |

aThe $S_1$ geometry optimization using RI-CC2/cc-pVDZ method leads to a first excited state energy of 4.77 eV.

From our calculations, it does not seem that a tropylium excited state with reasonable oscillator strength lies in the 5 eV energy region. The observed transition in the 5 eV region is broad and without any apparent vibrational structure (Fig. 2), which is indicative either of an excited state undergoing a fast non-radiative decay or of a very large change in geometry as compared to the ground state leading to spectral congestion. The first excited state optimization leads to a geometry close to that of the ground state, with an energy of 4.77 eV (RI-CC2/cc-pVDZ). To test the possibility of a conical intersection with another excited state at some distance from the equilibrium geometry, the first excited state has been optimized starting from a symmetry-broken geometry. The $D_{7h}$ symmetry structure is recovered with the optimization process, which indicates the presence of a local minimum on the excited state potential energy surface. Therefore, this state should exhibit sharp vibrational bands, as observed for the first excited state of the benzylium ion. Thus, it seems that the tropylium ion is not observed in the present experiment and may be difficult to observe in this UV spectral range, because of the low oscillator strengths of the transitions. Note that there is one state ($^1E'_1$ state in table II) that has a large oscillator strength. However, the geometry relaxation of this state has been performed and leads to a transition energy of 6.09 eV after optimization and thus is probably not responsible for the 5 eV absorption.

### 3. 1-phenylethyl cation (m/z 105)

The ground state optimized structure of the 1-phenylethyl cation has a phenyl-CH-CH$_3$ structure of C$_s$ symmetry (Fig. 5a). The phenyl-CH$_2$-CH$_2$ form is not a minimum in the ground state potential energy surface. Geometry optimization starting from the phenyl-CH$_2$- $CH_2$ form leads without barrier to the phenyl-CH-$CH_3$ form located at -0.77 eV in energy. The vertical transition energies given in Table III are calculated with a ground state optimized phenyl-CH-$CH_3$ geometry.

TABLE III. Vertical and adiabatic (bold italic) excited state transition energies (eV) and oscillator strengths of the 1-phenylethyl cation calculated at the CC2/cc-pVDZ level in $C_s$ symmetry. The experimental transition origins are also indicated (eV).

| | Vertical/*Adiabatic* transition energy | Oscillator strength | Exp. transition origin |
|---|---|---|---|
| $S_1$ ($^1A'$) | 3.22/***2.77***[a] | 0.032 | 2.61 |
| $S_2$ ($^1A'$) | 4.44 | 0.55 | 3.65 |
| $S_3$ ($^1A''$) | 5.19/***4.62*** | 0.000014 | |
| $S_4$ ($^1A''$) | 5.57 | 0.00004 | |

[a]*2.62* eV with the difference in zero-point energy correction.

The adiabatic $S_1$ ($1^1A'$) ← $S_0$ ($X^1A'$) transition energy, including the δZPE correction, has been calculated at 2.62 eV, which is in very good agreement with the observed band origin at 2.61 eV. It is interesting to note that in the $S_1$ state, the methyl group has rotated by 60° as compared to the ground state geometry. The stabilization energy due to this rotation can be calculated in freezing the methyl rotation in $C_s$ symmetry when optimizing $S_1$, and is 0.033 eV.

In the $S_2$ state ($2^1A'$ in $C_s$ symmetry) optimization process, the methyl group rotates out of the benzene plane along the $C_1$-$C_7$ axis, as seen in Figure 5. For an angle of 64°, the $S_1$ and $S_2$ states become degenerate and at 90° the $S_2$ state is the lowest state at 2.94 eV.

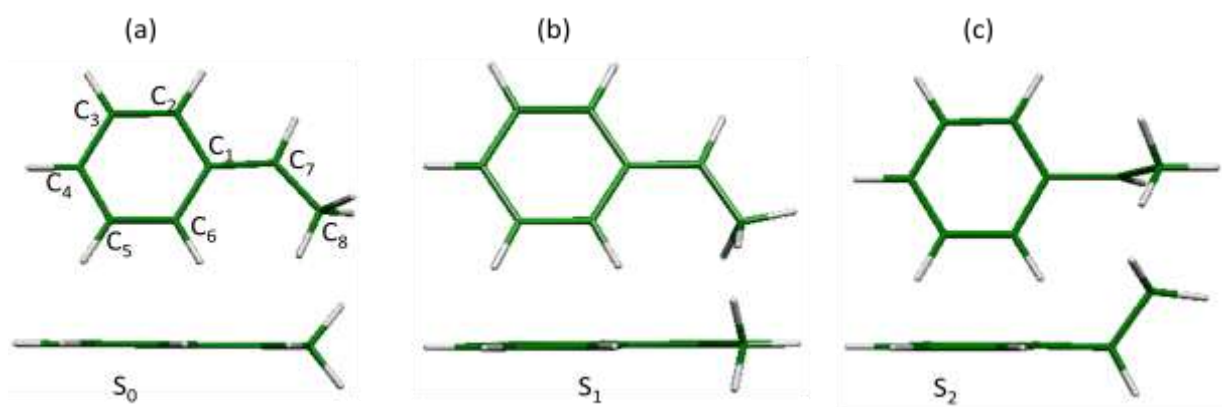

FIG. 5. (a), (b) and (c) Optimized geometries for the first electronic states ($S_0$, $S_1$ and $S_2$) of the 1-phenylethyl cation. In the $S_1$ state (b), the methyl group has rotated by 60° around the $C_7$-$C_8$ axis as compared to $S_0$ (a), but the molecule still keeps the benzene plane as plane of symmetry. In the $S_2$ state (c), the $CH_3$ group rotates around the $C_1$-$C_7$ axis and comes out of the benzene plane.

The first A" state calculated vertically at 5.19 eV can be optimized in $C_s$ symmetry, which results in an adiabatic transition energy at 4.62 eV. The change in geometry with respect to the ground state is mostly due to an increase of the $C_1$-$C_4$ distance in the aromatic ring from 2.80 Å to 3.026 Å as in $S_1$. Thus we would expect to see a progression of 400/500 cm$^{-1}$ as in the $S_1$ ($1^1A'$) ← $S_0$ ($X^1A'$) transition.

### 4. The methyl-tropylium (m/z 105)

As for the tropylium ion, the methyl-tropylium has been calculated at the CC2/cc-pVDZ level (Table IV). The first excited state does not exhibit a large change in geometry as compared to the ground state and has non-negligible oscillator strength (0.019). If present in the experiment, it should be characterized by well-defined vibrational structures with sharp peaks around 37 000 cm$^{-1}$.

TABLE IV. Vertical and adiabatic (bold italic) excited state transition energies (eV) of the methyl-tropylium cation calculated at the CC2/cc-pVDZ level.

|       | vertical/*adiabatic* transition energy | Oscillator strength |
|-------|----------------------------------------|---------------------|
| $S_1$ | 4.73/***4.60***                        | 0.019               |
| $S_2$ | 4.82                                   | 0.003               |
| $S_3$ | 6.29                                   | 0.544               |

## IV. Discussion
### A. *The first excited state in the visible region*
#### 1. Benzylium ion (m/z 91)

The spectrum recorded in the visible region for cold, bare benzylium ions is very similar to the one obtained for the argon complex.[47] This justifies that indeed the messenger technique does not perturb much the spectroscopic measurements and that the shift of the band origin can be correctly predicted:[59] the estimated value obtained from the argon tagging experiment (19 082 ± 20 cm$^{-1}$) is in remarkable agreement with our measurement of the $S_1$ band origin at 19 089 ± 10 cm$^{-1}$. The visible spectrum of the argon complex shows a main vibrational progression built on $\nu_{13}$, a ring deformation mode that reflects the change in geometry between the ground and excited states.[47] The same progression is observed here and in addition, three other progressions involving $\nu_{13}$ are observed starting on vibrational modes at 1176, 1236 and 1350 cm$^{-1}$. These modes can be tentatively assigned on the basis of calculated frequencies and simulated spectrum (see the simulated spectrum in Figure SI2 of the Supplementary Material). The 1176 cm$^{-1}$ band is assigned to the $9_0^1$ level, the $\nu_9$ mode being Franck-Condon active in agreement with Dryza et al.[36] The band at 1350 cm$^{-1}$ is assigned to the $30_0^2$ vibrational level being Franck-Condon active: the $\nu_{30}$ mode corresponds to the out of plane bending of the CH$_2$ group. The 1236 cm$^{-1}$ band is tentatively assigned to the $8_0^1$ vibrational level calculated at 1203 cm$^{-1}$. It should be noted that these excited state frequencies are close to the ground state frequencies reported in previous studies ($13_1^0$ at 525 cm$^{-1}$, $9_1^0$ at 1187 cm$^{-1}$, $8_1^0$ at 1355 cm$^{-1}$) (see table SI1 in the Supplementary Material for a more complete comparison of calculated and experimental frequencies).[47,60–62]

All the progressions show the same band splittings: all the bands containing $\nu_{13}$ are split, by 14 cm$^{-1}$ for the band involving one quantum in $\nu_{13}$, 21 cm$^{-1}$ for the band involving two quanta in $\nu_{13}$, and 37 cm$^{-1}$ for the band involving three $\nu_{13}$ quanta. As a consequence, the value of the $\nu_{13}$ vibration is difficult to evaluate, and is estimated to be 504 ± 8 cm$^{-1}$, a value which is slightly smaller than that obtained in previous experiments (Ar tagging or matrix).

The band splittings are not due to hot bands: no hot bands have been observed to the red of the band origin, and the simulated spectrum shows only one weak hot band ($36^1_1$ in the simulated spectrum at 100 K) that does not seem to be present in the experimental spectrum (see Figure SI4 in the Supplementary Material). The splittings observed in the experimental spectra are most probably due to Fermi resonances. In our calculations as well as in previous ones,[47] there is no vibrational mode close to the $\nu_{13}$ mode, but the combination of two low out-of-plane modes could be responsible for Fermi resonances. Indeed, a combination of the lowest mode calculated at 185 cm$^{-1}$ with another out-of-plane mode calculated at 313 cm$^{-1}$ leads to 498 cm$^{-1}$, not far from the value of the $\nu_{13}$ mode.

The bands that do not involve $\nu_{13}$ quanta in the spectrum of cold benzylium are narrower (FWHM = 12 ± 2 cm$^{-1}$ at the $S_1$ origin and on the $9^1_0$ and $8^1_0$ bands) than the bands of the Bz$^+$-Ar complex (FWHM~60 cm$^{-1}$ at the $S_1$ origin) indicating probably a participation of van der Waals low frequency modes in the profiles recorded in the Ar tagged experiment.

### 2. 1-phenylethyl cation (m/z 105)

The spectrum of 1-phenylethyl cation in its first excited state is quite similar to the benzylium one, except that low frequency vibrations (c.a. 50 cm$^{-1}$) are superimposed on the $\nu_{13}$ vibrational progression. The calculations have been performed for the phenyl-CH-CH$_3$ form because the phenyl-CH$_2$-CH$_2$ form has no minimum in the ground state energy surface and the calculated electronic transition is 2.62 eV, in very good agreement with the experimental value of 2.61 eV. This indicates that the ion structure has the CH-CH$_3$ form and not the CH$_2$-CH$_2$ form, which could be expected from the protonated phenylethylamine precursor after NH$_3$ loss. Since the calculation shows a rotation of the methyl group between $S_0$ and $S_1$, one expects to see a progression along the rotation coordinate. The lowest vibration calculated in $S_1$ at 50 cm$^{-1}$ is not a pure CH$_3$ rotation but it is linked to the C-CH-CH$_3$ bending angle. The second lowest frequency, at 116 cm$^{-1}$ in S$_1$ and 145 cm$^{-1}$ in S$_0$ corresponds to the methyl rotation, but the spectral simulation performed with PGOPHER[57] does not reproduce the observed spectrum and in particular fails to reproduce the low frequency progression (see figure SI3 in the Supplementary Material). However as seen in the inset of Figure 3, the spectrum looks similar to the benzylium spectrum if one superimposes a low frequency progression to the active $\nu_{13}$ mode. The low frequency progression can be modeled with a one dimensional hindered rotation model (Suzuki et al)[63] to reproduce the band pattern in the low energy part of the spectrum, using a potential $V(\phi)=V_3(1-\cos(3\phi))+V_6(1-\cos(6\phi))$, where $\phi$ is the torsional angle and $V_3$ and $V_6$ are the amplitudes of the $3\phi$ and of the $6\phi$ components of the potential. Since the CH$_3$ rotation is coupled to the CH$_3$ out-of-plane motion, the effective rotation constant should be smaller than in a molecule like cresol (B=5 cm$^{-1}$) and is considered as a free parameter. A reasonable fit is obtained using the following parameters: B=1.5 cm$^{-1}$, $V_3$"= -90 cm$^{-1}$, $V_6$" = -40 cm$^{-1}$ for the ground state and $V_3$' = 110 cm$^{-1}$, $V_6$' = -10 cm$^{-1}$ for the excited state. The barriers obtained with the fit are smaller than the ab initio calculations. As seen in the figure SI3 of the Supplementary Material, the agreement between the fit using the hindered rotation model and the experimental spectrum is fairly good implying that the methyl torsion is rotated by $\pi/3$ in the excited state as compared to the ground state (see supplementary information).

### B. Higher excited states in the UV region
#### 1. Benzylium ion (m/z 91)

The second electronic state of the benzylium ion starting at around 3.95 eV may be assigned to the second excited state (1$^1$A$_1$), calculated at 4.43 eV in the C$_{2v}$ symmetry (without δZPE correction). But when the C$_{2v}$ symmetry is released, the CH$_2$ group rotates without barrier and gets perpendicular to the aromatic ring, with its energy decreasing to 2.44 eV. With such a geometry change, the 0-0 transition is not expected to be observed due to vanishing Franck-Condon factors. Broad bands are observed in the 4 eV region (Fig. 2), which means that the potential energy surface is rather flat in the Franck-Condon region or has a kind of local minimum in the planar configuration. The first part of the spectrum shows broadened vibronic bands, as the spectrum recorded in neon matrices[16] but the width is larger and the spectrum seems to be blueshifted as compared to the matrix spectrum. The $CH_2$ rotation may be blocked in the condensed phase, which would lead to a sharper spectrum.

Around 5 eV, an intense very broad band is observed. It is not easily assigned to the $C_{2v}$ benzylium isomer. Indeed, the $S_3$ and $S_4$ states calculated at 5 and 5.4 eV have very small oscillator strengths. This band is large with no apparent vibrational progressions, which indicates either a fast non-radiative decay and/or a very large change in geometry. However one should keep in mind that the intensity in a photo-fragmentation spectrum depends on several factors that are not controlled, such as the fragmentation efficiency, so that the contribution of $S_3$ and $S_4$ states cannot be completely ruled out. It is also possible to assign this band to higher excited states of benzylium not calculated at the ground state geometry, which would undergo a strong change in geometry that could open Franck-Condon activity in this 5 eV region.

Based on the calculated electronic transitions in the 5 eV spectral range, it does not seem that the tropylium ion would contribute to the experimental spectrum (no allowed calculated electronic transitions in this spectral range). The first tropylium excited state that has a large oscillator strength is the $^1E'_1$ state, but the geometry relaxation of this state has been performed (TD-DFT in $C_s$ symmetry) and leads to an energy of 6.09 eV after optimization and thus is probably not responsible for the 5 eV absorption.

The presence of other linear isomers that could absorb in the 5 eV region in the trap cannot be excluded.

### 2. 1-phenylethyl cation (m/z 105)

As for benzylium ions, two very broad transitions are observed in the UV region, centered around 4 eV (33 000 cm$^{-1}$) and 4.8 eV (39 000 cm$^{-1}$). The second electronic transition starts at 3.65 eV with some vibrational progressions, which is not the case for the third transition at 4.8 eV.

In the 4 eV band, the fragmentation pathways are changing as a function of the excitation energy: at low energy, near the band origin (3.65 eV, 29 500 cm$^{-1}$), fragmentation leads preferentially to the C$_2$H$_4$ loss, while at higher energies (above 34 000 cm$^{-1}$, 4.21 eV), fragmentation mainly leads to C$_2$H$_2$ loss. We can assign this state to the second excited state of A' symmetry, calculated vertically at 4.4 eV, which has its equilibrium geometry in a structure where the C$_7$-C$_8$ bond is almost perpendicular to the phenyl ring (the relaxation from the planar to the perpendicular structure is barrier free). Thus, for this state, the Franck-Condon region should be far from the 0-0 transition and vibrational congestions should lead to absence of sharp vibrational structures. Besides, fast dynamical processes may be induced by conical intersections with other states: in going from the vertical structure (4.4 eV) to its optimized structure (around 2.9 eV), there is a crossing with the 1A' state. This is again very similar to the benzylium case.

The last band observed at 39 000 cm$^{-1}$ (4.8 eV) is tentatively assigned to the 1A" state calculated vertically at 5.19 eV and adiabatically at 4.62 eV. But as in the benzylium case, the calculated oscillator strength for this state is very low, whereas the band is relatively intense. However, once again, the intensity in a photo-fragmentation spectrum depends on the fragmentation efficiency that is not known.

Alternatively, this band could be assigned to the first excited state calculated for the methyl tropylium cation. But the calculated geometry change for this state with respect to the ground state is rather small, and therefore we would have expected a well-resolved vibronic spectrum. The question remains open, as in the case of the benzylium ion where the 5 eV band is not clearly assigned.

## V. Conclusion

The benzylium and 1-phenylethyl cations have been produced by collision-induced fragmentation of protonated benzylamine and protonated phenylethylamine from an electrospray ion source. The electronic spectra of benzylium and 1-phenylethyl cations were obtained via photo-fragment spectroscopy in an energy range from 2 eV to 5.5 eV. Both spectra show three major bands: the first one, in the visible, presenting sharp vibrational progressions, the second one still showing broadened vibronic bands while the third band is unstructured.

The visible spectrum of the benzylium cation ($S_1 \leftarrow S_0$ transition) has an origin at 19 089 cm$^{-1}$ (2.36 eV) and shows a main vibrational progressions built on the $\nu_{13}$ mode (around 504 cm$^{-1}$), and satellite progressions of $\nu_{13}$ in combination with totally symmetric $\nu_9$ and $\nu_8$ vibrational modes. Fermi resonances cause splittings of the bands involving the $\nu_{13}$ mode by a few tens of wavenumbers, probably due to a combination band of two out-of-plane vibrations close to $\nu_{13}$ mode. The visible spectrum of 1-phenylethyl cation shows a weak band origin and a vibrational progression on a mode of c.a. 400-500 cm$^{-1}$ that looks similar to the progression of the benzylium cation, on which is superimposed a low frequency progression reminiscent of a methyl hindered rotation.

The second electronic state, around 4 eV, shows a broadened vibrational progression in the two ions. This corresponds to a large geometry change as indicated by the RI-CC2 optimizations that give a 90° rotation of the methylidene group for the $S_2$ state of benzylium ion and an out-of-plane torsion around the $C_1$-$C_7$ axis for the $S_2$ state of the 1-phenylethyl cation.

The third electronic state (around 5 eV) is even broader, without any vibrational progression. The absence of vibronic structure can be interpreted as resulting from a fast non-radiative decay or from a very large change in geometry as compared to the ground state. However, the calculations performed in this study do not allow a clear assignment of the third band either to the benzylium and 1-phenylethyl cation or to their isomeric forms tropylium or methyl-tropylium ions. More accurate methods may provide insight to these excited states and their fragmentation pathways.

## Supplementary Material

"See Supplementary Material Document No.

for:

- Photo-fragmentation mass spectrum of the benzylium cation at 20 623 cm$^{-1}$ (2.56 eV)
- Comparison between the experimental and simulated spectra for benzylium cation in S$_1$
- Low energy region of the photo-fragmentation spectrum of the phenylethyl cation
- Zoom of the first two bands of benzylium (S$_1$) with the simulation
- Table of Comparison between calculated and observed frequencies

## Acknowledgments

This works was supported by the Aix-Marseille Université, the CLUPS of Université Paris–Sud 11, the ANR Research Grant (ANR2010BLANC040501). We acknowledge the use of the computing facility cluster GMPCS of the LUMAT federation (FR LUMAT 2764). Stéphane Coussan and Gaël Roussin are thanked for their help in setting up the experiment.

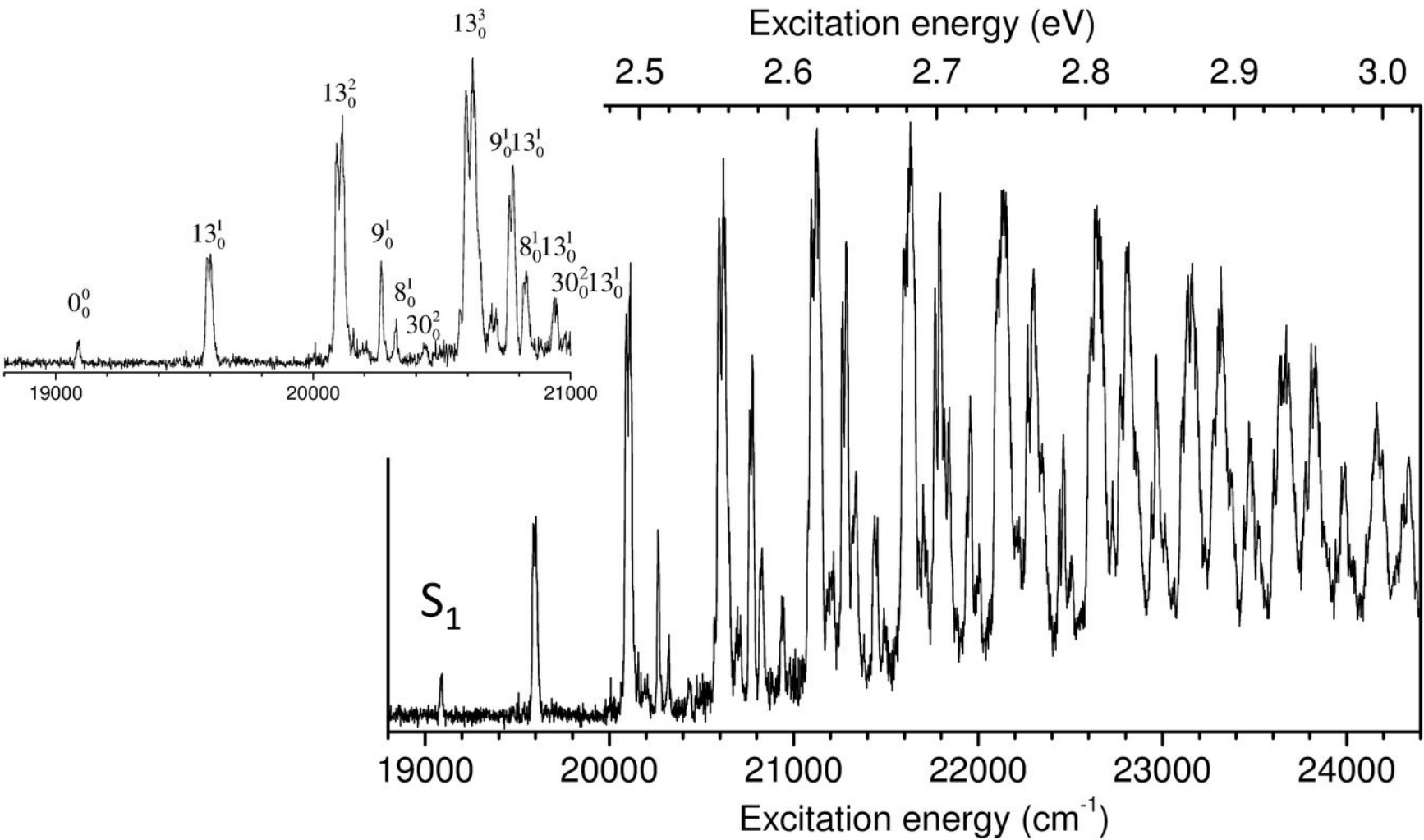

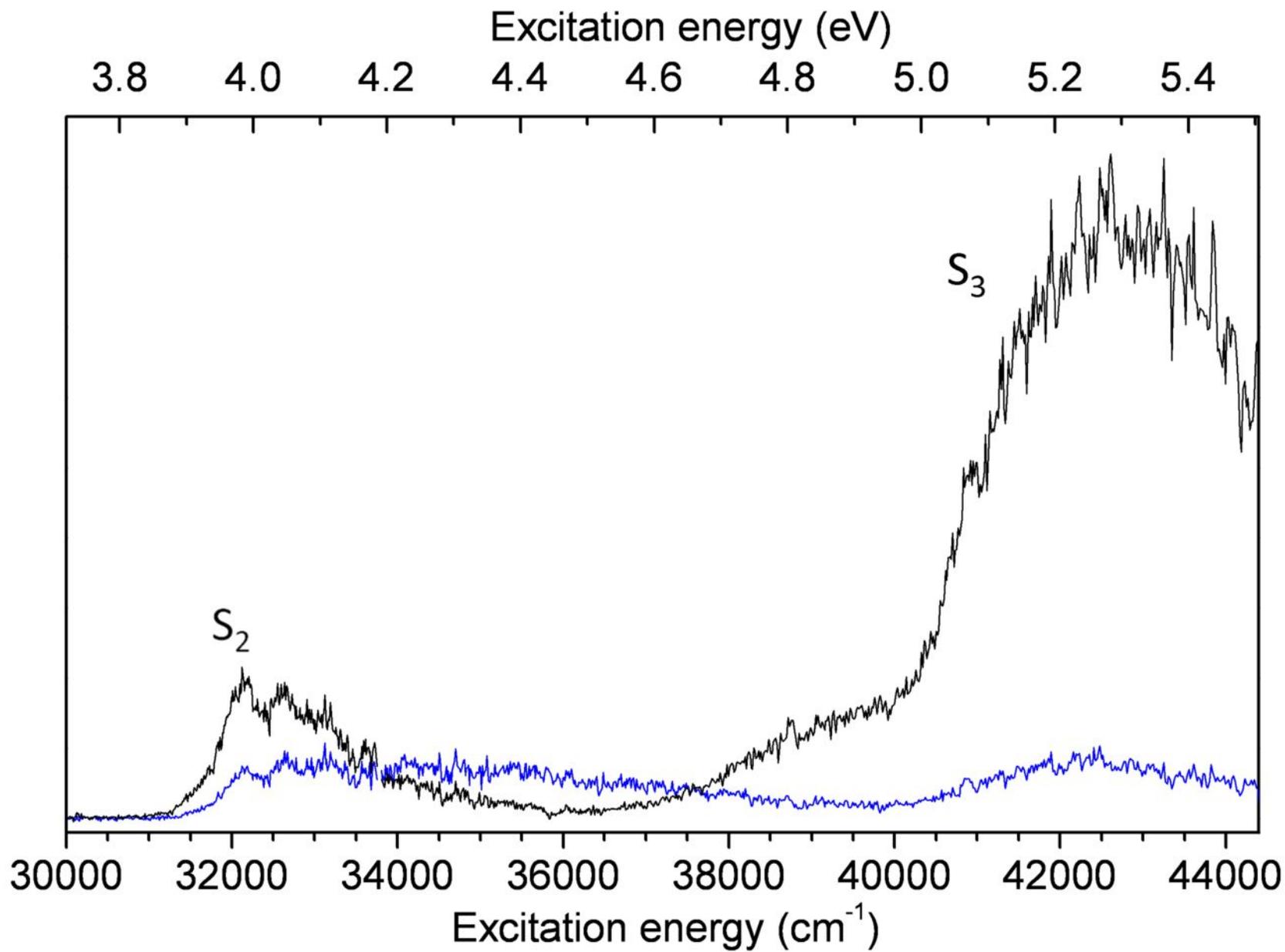

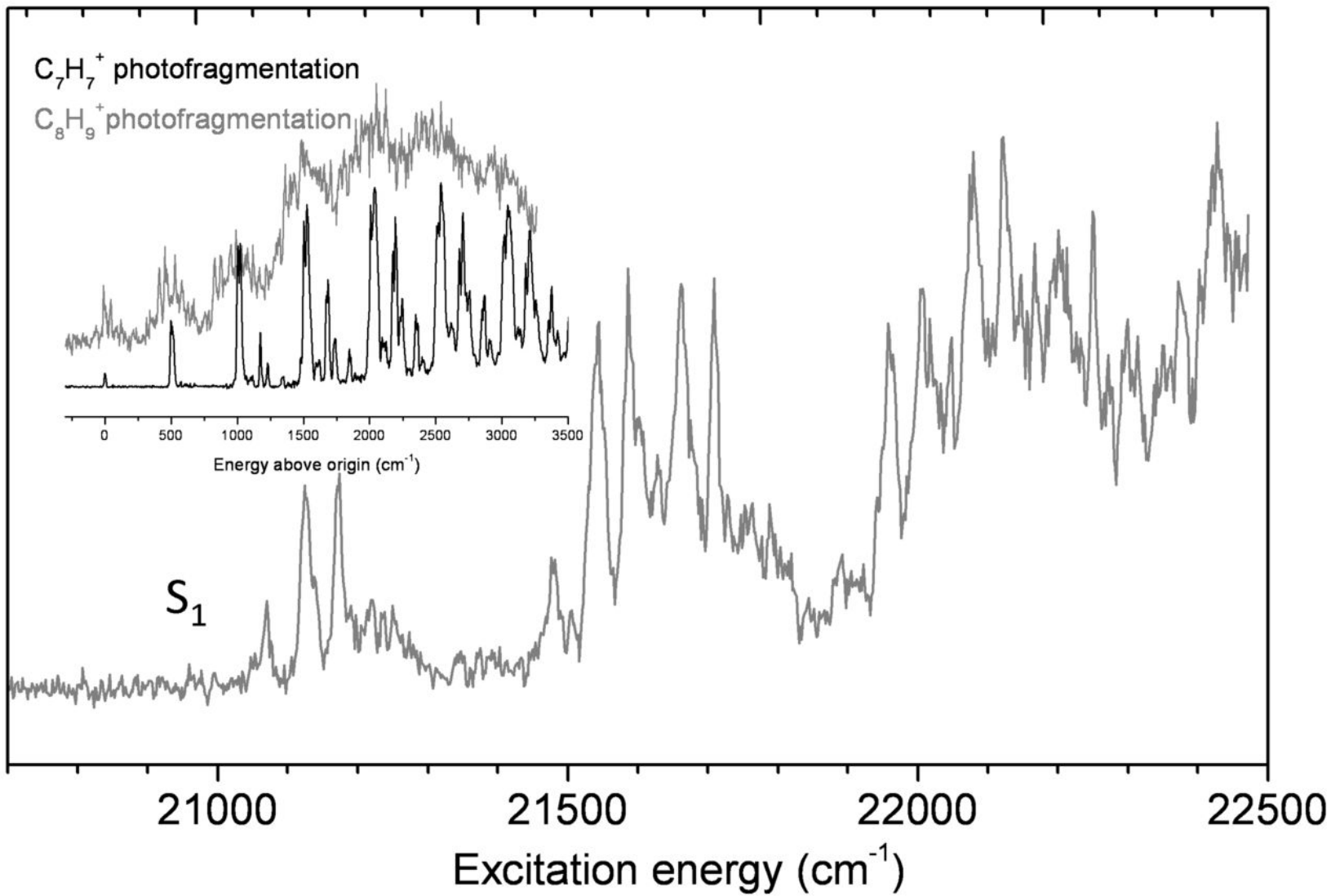

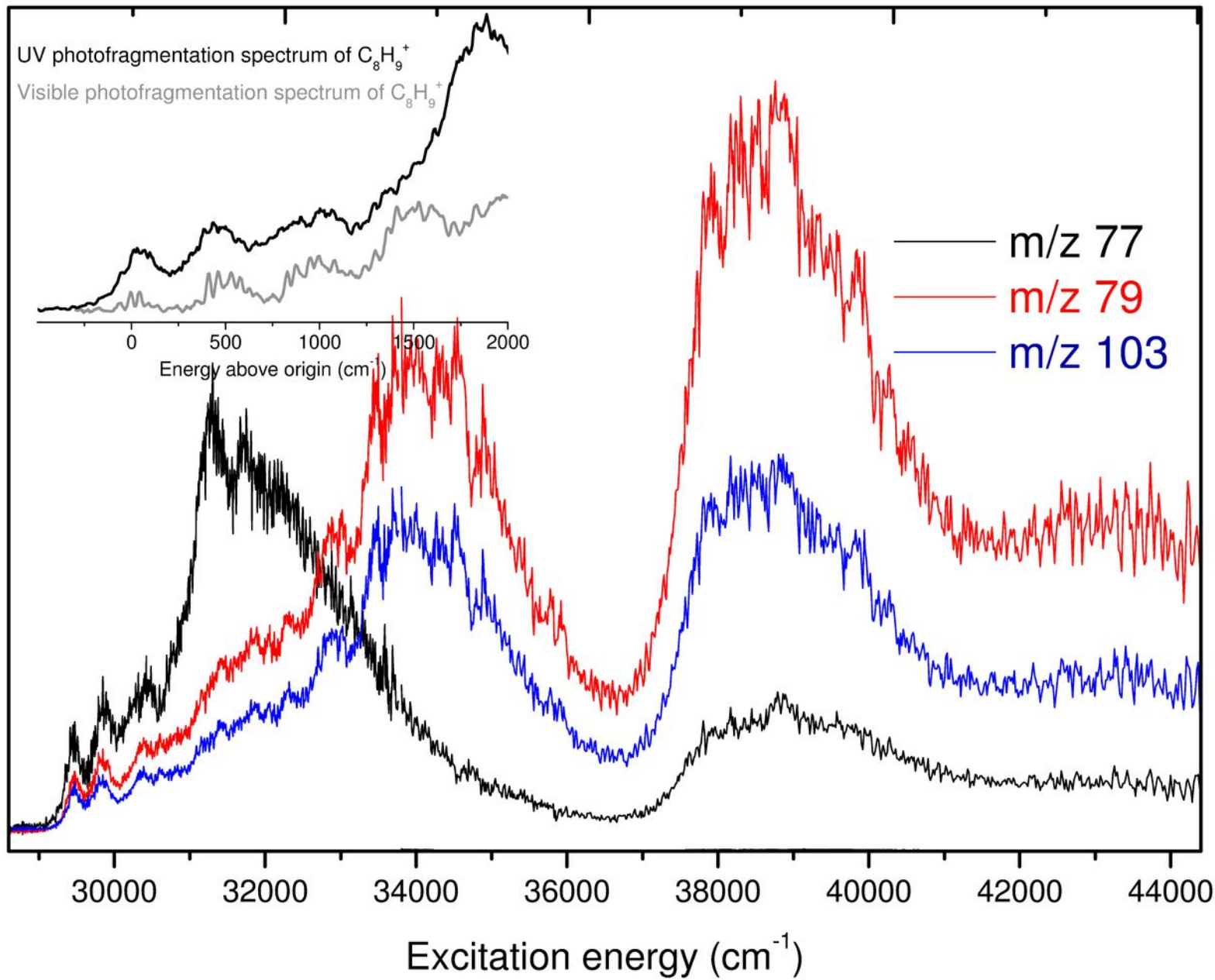

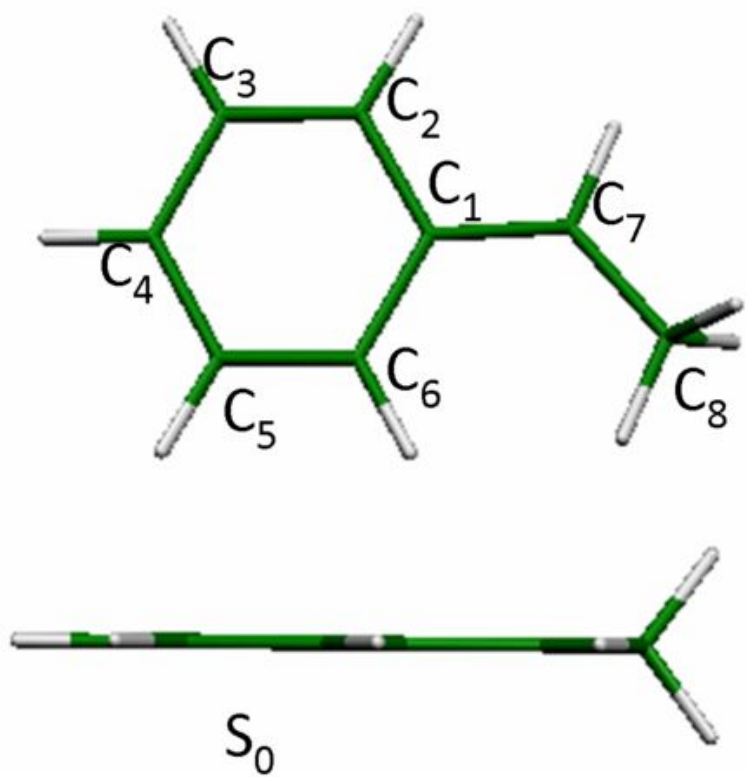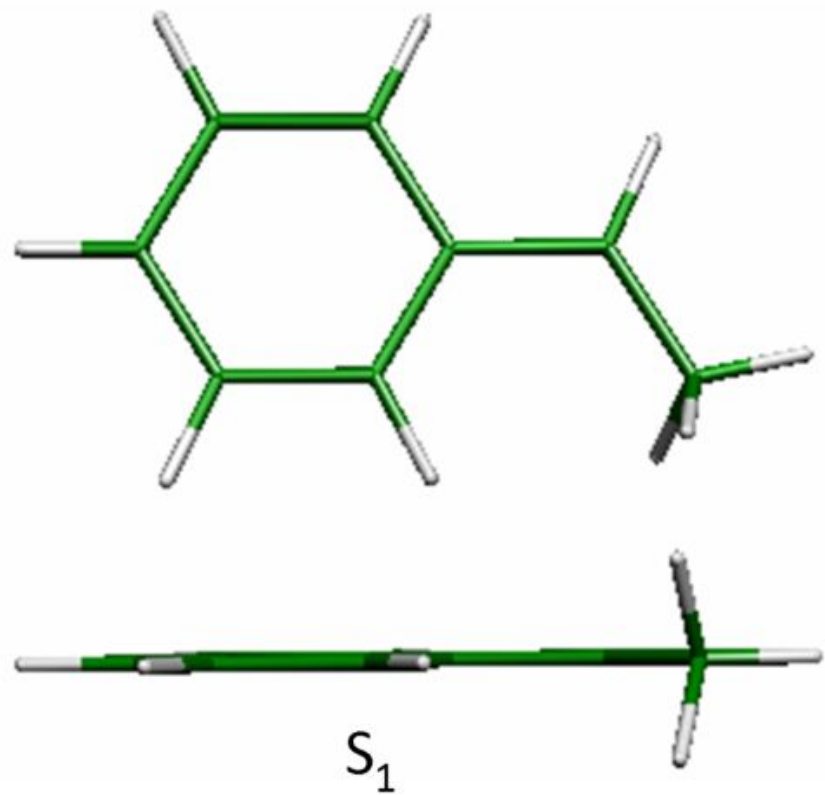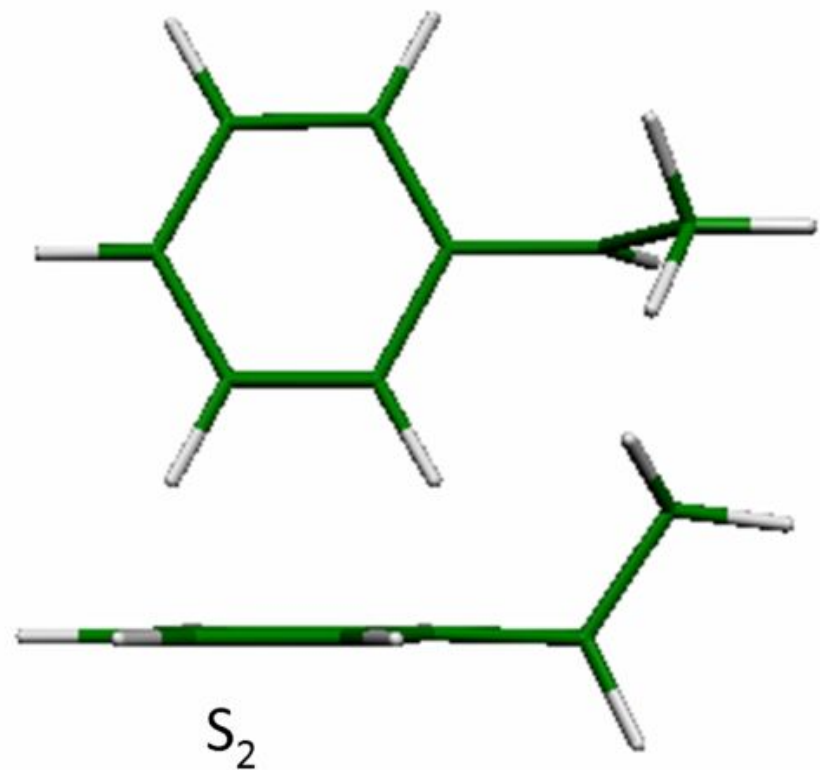